\documentclass[twocolumn,aps,prl,showpacs]{revtex4}

\usepackage[latin1]{inputenc}
\usepackage{amsmath}

\begin{document}
\title{Reply to Fahmi and Golshani's comment on ``Quantum key distribution in the Holevo limit''}
\author{Ad\'{a}n Cabello}
\email{adan@us.es}
\affiliation{Departamento de F\'{\i}sica Aplicada II,
Universidad de Sevilla, E-41012 Sevilla, Spain}

\date{\today}


\begin{abstract}
As Fahmi and Golshani correctly point out, a protocol introduced in
A. Cabello, Phys. Rev. Lett. {\bf 85}, 5635 (2000), to show that a
quantum key distribution protocol can have efficiency one (i.e., can
achieve the Holevo limit), does indeed not have efficiency one. The
corrected protocol, introduced in A. Cabello, Recent. Res. Devel.
Physics {\bf 2}, 249 (2001), is reproduced here.
\end{abstract}




\maketitle


As Fahmi and Golshani correctly point out in the preceding Comment
\cite{FG07}, a protocol introduced in \cite{Cabello00} to show that
a quantum key distribution protocol can have efficiency one (i.e.,
can achieve the Holevo limit), where efficiency is defined as the
number of secret bits per transmitted bit plus qubit, does indeed
not have efficiency one. This error was already corrected in
\cite{Cabello01,Cabello02}. For completeness' sake, the corrected
protocol introduced in \cite{Cabello01,Cabello02}, with efficiency
one, is reproduced here.

Suppose that the quantum channel is composed of two qubits ($1$ and
$2$) prepared with equal probabilities in one of four orthogonal
pure states:
\begin{mathletters}
\begin{subequations}
\begin{align}
|\eta_0\rangle = {1 \over \sqrt3} \left(|00\rangle +
|01\rangle + |10\rangle\right), \label{eta0} \\
|\eta_1\rangle = {1 \over \sqrt3} \left(|00\rangle -
|01\rangle + |11\rangle\right), \\
|\eta_2\rangle = {1 \over \sqrt3} \left(|00\rangle -
|10\rangle - |11\rangle\right), \\
|\eta_3\rangle = {1 \over \sqrt3} \left(|01\rangle - |10\rangle +
|11\rangle\right). \label{eta3}
\end{align}
\end{subequations}
\end{mathletters}
Alice sends both qubits to Bob. Eve cannot access qubit~$2$ while
she still holds qubit~$1$. Each pair of qubits encodes $2$ bits of
the key (for instance, ``00'' if the state is $|\eta_0\rangle$,
``01'' if the state is $|\eta_1\rangle$, ``10'' if the state is
$|\eta_2\rangle$, and ``11'' if the state is $|\eta_3\rangle$).
Since the four states (\ref{eta0})--(\ref{eta3}) are orthogonal, Bob
can unambiguously discriminate between them and identify which is
the one sent by Alice.

As can be easily checked, the revised protocol does not only satisfy
Mor's requirements to prevent cloning when Eve has a one-by-one
access to the qubits (namely, that the reduced states of the first
subsystem must be non-orthogonal and non-identical, and the reduced
states of the second subsystem must be non-orthogonal \cite{Mor98})
for {\em any} two states chosen from (\ref{eta0})--(\ref{eta3}), but
is also secure against the double C-NOT eavesdropping strategy
proposed by Fahmi and Golshani in \cite{FG07}.



\end{document}